\documentclass[lettersize,journal]{IEEEtran}
\usepackage{amsmath,amsfonts}
\usepackage{algorithmic}
\usepackage{algorithm}
\usepackage{array}
\usepackage[caption=false,font=normalsize,labelfont=sf,textfont=sf]{subfig}
\usepackage{textcomp}
\usepackage{stfloats}
\usepackage{url}
\usepackage{verbatim}
\usepackage{graphicx}
\usepackage{cite}

\usepackage{makecell}
\usepackage{textgreek}
\usepackage{graphicx} 
\graphicspath{{figures/}} 

\begin{document}

\title{Whole-body PET image denoising for reduced acquisition time}

\author{Ivan Kruzhilov, Stepan Kudin, Luka Vetoshkin, Elena Sokolova, Vladimir Kokh}



\maketitle

\begin{abstract}
This paper evaluates the performance of supervised and unsupervised deep learning models for denoising positron emission tomography (PET) images in the presence of reduced acquisition times. Our experiments consider 212 studies (56908 images), and evaluate the models using 2D (RMSE, SSIM) and 3D (SUVpeak and SUVmax error for the regions of interest) metrics. It was shown that, in contrast to previous studies, supervised models (ResNet, Unet, SwinIR) outperform unsupervised models (pix2pix GAN and CycleGAN with ResNet backbone and various auxiliary losses) in the reconstruction of 2D PET images. Moreover, a hybrid approach of supervised CycleGAN shows the best results in SUVmax estimation for denoised images, and the SUVmax estimation error for denoised images is comparable with the PET reproducibility error.      
\end{abstract}


\section{Introduction}

\begin{figure*}[!t]
\centering
\subfloat[]{\includegraphics[width=1.8in]{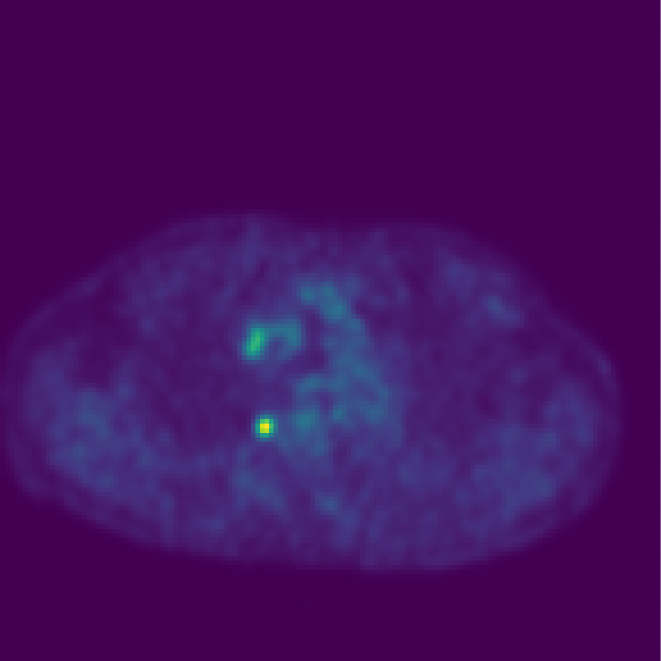}%
\label{fig_first_case}}
\hfil
\subfloat[]{\includegraphics[width=1.8in]{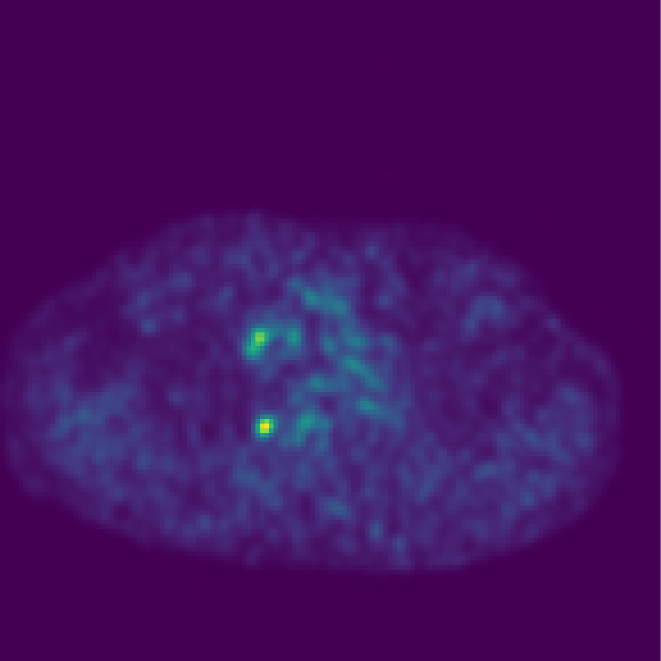}%
\label{fig_second_case}}
\hfil
\subfloat[]{\includegraphics[width=1.8in]{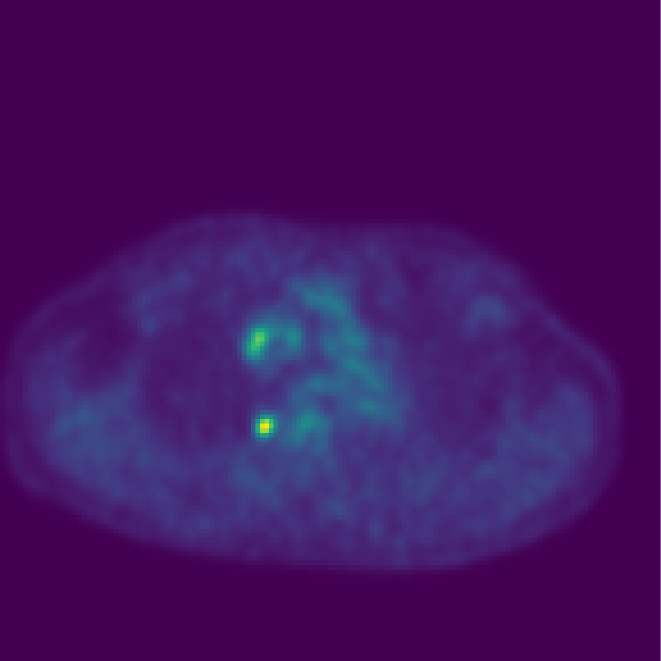}%
\label{fig_third_case}}
\caption{Low-time PET reconstruction. (a) Full-time PET (90 sec). (b) Low-time PET (30 sec). (c) Reconstructed PET (by Transformer)}
\label{fig_example1}
\end{figure*}

\IEEEPARstart{P}ET is a molecular imaging technique that produces a 3-dimensional radiotracer distribution map representing properties of biologic tissues, such as metabolic activity. Many patients undergo more than one PET/CT scan per year. The higher the injected activity, the less noise in the reconstructed images and the more radioactive exposure for a patient.

Deep learning methods may reduce injected activity or acquisition time by utilizing low-dose (LD) / low-time (LT) and full-dose (FD) / full-time (FT) images (\ref{fig_example1} to train models that can predict standard-dose images from LD / LT inputs. A reduced acquisition time positively impacts the patient’s comfort or the scanner’s throughput, which enables more patients to be scanned daily, lowering costs.

The study \cite{schaefferkoetter2019low} demonstrates that the statistical decimation of FD scans can accurately emulate clinical LD PET scans. This fact allows modelling reduced acquisition time instead of LD injection and vice versa because their statistical properties are equivalent.

The drawback of the recent studies is the need for more comparison between a broad group of methods on a level playing field, especially between supervised and unsupervised methods. Moreover, as methods are tested on different data sets and for various PET time frames, comparing them is complicated. Furthermore, studies differ in the metrics used for denoising quality assessment e.g., some \cite{yang2022quasi, jang2022spach, hu2022transem, spuhler2020full} evaluate only image similarity metrics like RMSE but do not take into account SUV characteristics.

Our study aims to overcome the drawbacks by finding the best backbone and model and comparing the performance of supervised and unsupervised methods for PET denoising. We tested supervised ResNet, Unet, and SwinIR \cite{liang2021swinir} and unsupervised pix2pix GAN and CycleGAN with identity and image prior losses \cite{yang2022quasi}. We decided to limit our study to 2D networks, though 3D or 2.5D networks could improve some metrics \cite{lu2019investigation}. We reconstructed FT 90 sec/bed position PET from LT PET 30 and 60 sec/bed position. 

We use the following metrics to measure the performance of the models: similarity index (SSIM), root mean square error (RMSE), R2, median and interquartile range (IQR). SSIM and RMSE measure the similarity of two 2D images - original and denoised. R2, median and IQR values describe the correlation between the tumor's standardized uptake values (SUV) characteristics (peak, max) measured on both original and denoised PET. SUV median and IQR in this study correspond to the same values of the Bland-Alman plot for the original and denoised PET. They show whether the generated images provide the same diagnostic information as the original ones.

We did not conduct a proper visual inspection of the enhanced image quality as the studies \cite{sanaat2021deep, weyts2022artificial, bonardel2022clinical} did but demonstrated our results for the renginologist community, which approved the high quality of our reconstructions.

 
\subsubsection*{\bf The main contribution of the article are}
\begin{itemize}
\item{The comparison of models with different backbones and losses trained and tested in the same conditions. Our comprehensive experiments showed the better performance (RMSE and SSIM) of the supervised methods over unsupervised ones for PET denoising though these results are not in line with some previous studies. At the same time, the hybrid supervised CycleGAN model has the smallest SUVmax (for the region of interest) estimation error.}
\item{First SwinIR application for PET denoising.}
\item{In previous studies \cite{sanaat2020projection, sanaat2021deep, weyts2022artificial, katsari2021artificial} professional renginologists did the malignant tumor segmentation. The disadvantages of this approach are the need for more qualified specialists, the cost and time of manual segmentation, and the inconsistency of results. In our work we segmented tumors automatically by nnU-Net, which makes our SUV error estimation pipeline reproducible.}
\end{itemize} 

The structure of the article is as follows - \ref{section_related_works} section is a review of the PET denoising methods; 
\ref{section_dataset} section gives information about the data set used for training and testing; \ref{section_methodology} section is about the models to be tested in the study and the PET denoising quality metric; \ref{section:neural_network} section contains details of the neural networks and training process; \ref{section_results} section reveals the results of our research; \ref{section_clinical_discussion} section is a comparison of the SUV error achieved in our study and SUV measurements reproducibility.

\section{Related works}
\label{section_related_works}
The latest (as of January 2023) review on the low-count PET reconstruction are \cite{chen2022image} and \cite{liu2021artificial}. The early studies of PET denoising treated specific parts of the human body, like the brain and lungs, and used small-size data sets producing low-quality reconstructions. For example, \cite{gong2018pet} utilized pretrained VGG19 and perception loss for supervised denoising lungs and brains. One could find the comprehensive overview of methods before 2020 in \cite{sanaat2020projection}. Table \ref{table:review} represents the summary of the later works on PET denoising, * - the authors did not specify what kind of SUV - mean, peak, or max they estimated.

\begin{table}[!t]
\caption{Summarization of low-count PET denoising methods. \label{tab:review}}
\label{table:review}
\centering
\begin{tabular}{|c||c||c||c||c||c|}
\hline
\thead{Ref.\\year} & \thead{Input\\ data,\\number\\of\\studies} &\thead{Network\\archi-\\tecture} &\thead{Low\\count,\\
full\\count} &\thead{Scanner\\model} &\thead{Metrics} \\
\hline
\makecell{\cite{sanaat2020projection}\\2020} & \makecell{Brain\\sino-\\grams,\\140} &\makecell{3D\\Unet} &\makecell{--,\\20 min} & \makecell{Siemens\\Biograph\\mCT} & \makecell{RMSE,\\SSIM,\\PSNR,\\SUV\textsuperscript{*}\\bias\\STD} \\
\hline
\makecell{\cite{sanaat2021deep}\\2021} & \makecell{Whole\\body,\\100} &\makecell{Cycle\\GAN,\\ResNet} &\makecell{3 min,\\27 min} &\makecell{Siemens\\Biograph\\mCT} & \makecell{RMSE\\SSIM\\PSNR\\Visual\\eval \\SUV\textsuperscript{*}\\bias\\STD\\R2}\\
\hline
\makecell{\cite{sanaei2021does}\\2021} & \makecell{brain\\head \\neck,\\140} &\makecell{Nifty\\Net\\High\\ResNet} &\makecell{2-4\%\\6\%,\\100\%} & \makecell{Siemens\\Biograph\\6} & \makecell{\\SSIM,\\PSNR,\\RMSE,\\SUVmean\\bias\\}\\
\hline 
\makecell{\cite{weyts2022artificial} \\ 2022} & \makecell{Whole\\body,\\195} &\makecell{Subtle\\PET\\2.5D\\Unet} &\makecell{45 sec,\\90 sec} &\makecell{VEREOS\\Philips\\Health\\care} &\makecell{SULmax\\SULpeak\\median\\IQR\\visual\\eval} \\
\hline
\makecell{\cite{bonardel2022clinical} \\ 2022} & \makecell{Whole\\body,\\100} &\makecell{Subtle\\PET\\2.5D\\Unet} &\makecell{33\%,\\50\%,\\100\%} &\makecell{GE \\ Discovery\\MI 4\\ 710,\\IQ4} &\makecell{\textDelta CRS\\ \textDelta BV\\ \textDelta CNR\\SUVmax \\visual\\eval} \\
\hline
\makecell{\cite{yang2022quasi}\\2022} & \makecell{Brain\\15000\\images} &\makecell{Cycle\\GAN\\super\\vised,\\quasi\\super\\vised} &--- &\makecell{---} & \makecell{RMSE,\\SSIM,\\PSNR}\\
\hline
\makecell{\cite{jang2022spach}\\2022} & \makecell{Whole\\body\\112} &\makecell{Unet,\\ Swin,\\ Restormer} &\makecell{25\%,\\ 100\%} &\makecell{GE DMI} &\makecell{SSIM,\\PSNR,\\CNR}\\
\hline
\makecell{Our\\2023} &\makecell{Whole\\body,\\ 212} &\makecell{SwinIR,\\ResNet,\\Unet,\\Cycle\\GAN\\image\\prior,\\pix2pix\\GAN} &\makecell{30 sec,\\ 60 sec, \\ 90 sec} &\makecell{GE, \\Discovery\\ 710} &\makecell{RMSE,\\SSIM,\\SUVmax,\\SUVpeak\\median\\bias,\\IQR}\\
\hline 
\end{tabular}
\end{table}

The Table \ref{table:review} demonstrates the usage of both supervised and unsupervised methods. CycleGAN is the most popular unsupervised model for the PET denoising. The article \cite{lei2019whole} was the first applied CycleGAN model for whole-body PET denoising. \cite{cui2019pet} also used unsupervised learning. CycleGAN showed better performance over Unet and Unet GAN in peak signal-to-noise ratio (PSNR) for all human body parts. \cite{sanaat2021deep} also utilized CycleGAN architecture and demonstrated its performance over ResNet, both trained on 60 studies data set. The ResNet showed, in turn, better results than Unet, which coincides with the results of our experiments. 

The works \cite{lei2019whole, sanaat2021deep} do not reveal the CycleGAN backbone used in their studies; therefore, it remains to be seen if CycleGAN in \cite{lei2019whole, sanaat2021deep} achieved high performance due to the unsupervised scheme and the adversarial losses or because of difference in the backbone.

Unlike other works studying cancer \cite{yang2022quasi} applied CycleGAN for the for Alzheimer's syndrome analysis. CycleGAN model also finds its application in different medical image denoising problems like optical coherence tomography images
\cite{manakov2019noise} and low-dose X-ray CT \cite{kwon2021cycle, tang2019unpaired}. 3D CycleGAN framework with self-attention generates the FC PET image from LC PET with CT aid in the paper \cite{lei2020low}.

The PET denoising problem is very similar to the PET reconstruction from CT. \cite{chandrashekar2022deep} demonstrated that non-contrast CT alone could differentiate regions with different FDG uptake and simulate PET images. To predict three clinical outcomes, \cite{chandrashekar2022deep} constructed random forest models on the radiomic features using the simulated PET. The objective of this experiment was to compare predictive accuracy between the Cycle-GAN-simulated and FT PET. ROC AUC for simulated PET achieved to be comparable with ground truth PET - 0.59 vs. 0.60, 0.79 vs. 0.82, and 0.62 vs. 0.63. The study \cite{li2021novel} denoised CT images by a GAN with the reconstruction loss. The study demonstrated the advantage of using image gradient information as GAN conditional information.

The most popular supervised models (Table 1 in \cite{liu2021artificial}) for the PET denoising are ResNet (e.g. \cite{sanaat2021deep}) and Unet-style networks (e.g. \cite{sano2021denoising, schaefferkoetter2020convolutional}). The article \cite{sanaei2021does} used HighResNet which demonstrated that due to the stochastic nature of PET acquisition, any LD versions of the PET data would bear complementary/additional information regarding the underlying signal in the standard PET image. This complementary knowledge could improve a deep learning-based denoising framework and (as \cite{sanaei2021does} showed) enhance the quality of FD prediction - PSNR increased from 41.4 to 44.9 due to additional LD images.

Rather than directly outputting the denoised CT image, \cite{Luthra2021EformerEE} used transformer-based encoder-decoder to predict the residual value. The method proposed in \cite{Luthra2021EformerEE} achieved the best RMSE metric but failed to outperform CNN in SSIM.

Swin transformer was used in \cite{hu2022transem} for FD brain image reconstruction from LC sinograms. The article \cite{jang2022spach} proposed spatial and channel-wise encoder-decoder transformer - Spatch Transformer that demonstrated better denoising quality over Swin transformer, Restormer, and Unet for 25\% low-count PET. The authors experimented with different tracers: $\textsuperscript{18}$F-FDG, $\textsuperscript{18}$F-ACBC, $\textsuperscript{18}$FDCFPyL, and $\textsuperscript{68}$Ga-DOTATATE. The tracers $\textsuperscript{18}$FDCFPyL and $\textsuperscript{68}$Ga-DOTATATE were used for test only to evaluate the robustness of the models. 

SubtlePET\textsuperscript{TM} is a commercial product; its official site claims that "SubtlePET is an AI-powered software solution that denoises images conducted in 25\% of the original scan duration (e.g., one minute instead of four)". SubtlePET uses multi-slice 2.5D encoder-decoder U-Net \cite{weyts2022artificial} optimizing L1 norm and SSIM. The networks were trained with paired low- and high-count PET series from a wide range of patients and from a large variety of PET/CT and PET/MR devices (10 General Electric, 5 Siemens, and 2 Philips models). The training data included millions of paired image patches from hundreds of patient scans with multi-slice PET data and data augmentation.

The recent studies \cite{weyts2022artificial,katsari2021artificial,yang2022quasi} investigated FT 90 sec PET reconstruction from LT 30, 45, and 60-sec images using SubtlePET. The work \cite{weyts2022artificial} conducted a study on the efficiency of SubtlePET by comparing denoised LT 45 sec PET with FT 90 sec PET. The visual analysis revealed a high similarity between FT and reconstructed LT PET. SubtlePET detected 856 lesions for 162 (of 195) patients. Of these, 836 lesions were visualized in both original 90 sec PET and denoised 45 sec PET, resulting in a lesion concordance rate of 97.7\%.

The study \cite{bonardel2022clinical} examined the limits of the SubtlePET denoising algorithm applied to statistically reduced PET raw data from 3 different last-generation PET scanners compared to the regular acquisition in phantom (spheres) and patient. Enhanced images (PET 33\% + SubtlePET) had slightly increased noise compared to PET 100\% and could potentially lose information regarding lesion detectability. Regarding the patient data sets, the PET 100\% and PET 50\% + SubtlePET were qualitatively comparable. In this case, the SubtlePET algorithm was able to correctly recover the SUVmax values of the lesions and maintain a noise level equivalent to FT images.

PET denoising is an inverse problem; to estimate the reconstruction uncertainty \cite{cui2022pet} proposed Nouveau variational autoencoder-based model using quantile regression loss.

\section {Whole-body PET data set}
\label{section_dataset}
There are two ways \cite{chandrashekar2022deep} to simulate low-dose PET - shot-frame and decimation. The most common way of decimation is the simulation of a dose reduction by randomized subsampling of PET list-mode data. Another method of decimation is randomly sampling the data by a specific factor in each bin of the PET sinogram \cite{sanaat2020projection}.

Short time frames with the corrections taking a shorter amount of time into account will produce images with similar SUV uptake as the original one. We use this approach in our study collecting PET data with 30, 60 and 90 sec / bed position. All images were collected during the same acquisition session that differs, e.g., from \cite{sanaat2021deep, sanaat2020projection} were the LT images obtained through a separate fast PET acquisition corresponding to the FT scans.

As \cite{sanaat2021deep} emphasized "there are a number of fundamental differences between LD images generated through decimating the FD scan and LD images actually acquired separately by reducing the acquisition time or the injected activity. First, when the LD PET image is obtained from a separate acquisition, the underlying PET signal may be different between LD and FD PET images owing to the varying tracer kinetics of the radiotracer during the course of imaging. Moreover, potential patient motion between these two scans further adds to the complexity of FD PET estimation from the fast/LD PET scan."

The slice thickness is 3.75 mm. The following coefficient \eqref{suv_coeff} transfers initial data into SUV (MBq/kG). The studies are in anonymized *.dcm format, from which one can extract the patient weight, half-life, and total dose values and delay between the injection time and the scan start time \textDelta t.
\begin{equation}
\label{suv_coeff}
 SUVcoeff = \frac{2000*weight}{total{\_}dose}*0.5^{-\frac{\Delta t}{half{\_}life}}
\end{equation}

After a 6-h fasting period and blood glucose levels testing, patients were injected with 7 MBq/kG [18F]FDG intravenously.

The train subset contains 160 studies with 42656 images, validation and test data both consist of 26 studies and 7126 images respectively. The patient age lies between 21 and 84 years and two thirds of patients are between 49 and 71 years old. The median age is 61; 71\% of patients are women. The number of tumors detected is 74 in the test subset and 97 in the validation.

The data set we used in this study is the largest PET whole body data set used so far (taking into account that SubtlePET is a commercial product and the authors do not reveal all details of their algorithms).

\section{Methodology}
\label{section_methodology}
\subsection{Problem statement}
The study aims to assess the quality of the PET denoising for supervised and unsupervised models. 
Unet, ResNet, CycleGAN, pix2pix GAN, SwinIR transformer are models to be tested in this research. More details of the network architecture are in \ref{section:neural_network}. In this paper, we also aim to evaluate the impact of the image prior and identity losses on the CycleGAN training and study the supervised version of CycleGAN with reconstruction loss to gain profits from both CycleGAN architecture and the availability of paired data.

In the original CycleGAN paper \cite{zhu2017unpaired} identity mapping loss
\begin{align}
\label{ident_loss}
ident = \sum_{i=0}^{n} {|denoised(FT_{i}) - FT_{i}|_{L1}} +\nonumber \\|noised(LT_{i}) - LT_{i}|_{L1}
\end{align}
helps preserve the color of the input painting. The loss steers the network not to denoise the FT image and vice versa. \cite{park2021effect} claims that more weights for the cycle-consistency loss and identity loss made the CycleGAN model translate the blood-pool image close to the actual bone image. We will investigate the influence of identity loss on PET denoising.

CycleGAN is an unsupervised method. Therefore, its usage is beneficial if there is a lot of unpaired data in both domains. But getting paired data with different PET acquisition times is an ordinary task that could be done automatically without any additional action on a patient. The study \cite{yang2022quasi} showed that the use of the additional supervised reconstruction loss \eqref{dist_loss} in CycleGAN makes the training stable and considerable improves PSNR and SSIM
\begin{equation}
\label{dist_loss}
rec{\_}loss = \frac{1}{n}\sum_{i=0}^{n} {|denoised(LT_{i}) - FT_{i}|_{L1}}.
\end{equation}
We used supervised CycleGAN as an upper boundary for ISSIM and RMSE metrics that unsupervised CycleGAN with the image prior loss could achieve and also study it effect on SUVmax error.

We trained CycleGAN with identity. \cite{zhu2017unpaired} and image prior \cite{tang2019unpaired} losses in addition to adversarial and cycle consistency losses. The idea of image prior loss 
\begin{equation}
\label{img_prior_loss}
img{\_}prior = \sum_{i=0}^{n} {|denoised(LT_{i}) - LT_{i}|_{L1}}.
\end{equation}
is based on the assumption of similarity between LT noised and FT original PET slices. It performs a regularization over CycleGAN generators preventing them from generating denoised PET images very different from the original one.

Gaussian convolution is a baseline model for denoising. The parameters of the filter are optimized on the validation data set. All models shared the same learning schedule and parameters (with minor differences described in the next section) and therefore have the same level playing field and could be fairly compared.
Table \ref{table:methods} is a systematization of the methods and models used in the study.
\begin{table}[!t]
\caption{The studied models\label{tab:table1}}
\label{table:methods}
\centering
\begin{tabular}{|c||c||c||c|}
\hline
\thead{type of model} & \thead{GANs \\ unsupervised} &\thead{supervised} &\thead{GANs + \\ supervised}\\
\hline
transformer & -- &SwinIR &--\\
\hline \makecell{Convolu\\tional \\ (Unet, ResNet)} & CycleGAN & \makecell{Unet, \\ ResNet\\ + \\decoder } &\makecell{pix2pix \\ GAN, \\CycleGAN \\superv}\\
\hline
\end{tabular}
\end{table} 

\subsection{Denoising quality assessment}

Two metric types describe the quality of denoising: a similarity of 2D PET images and concordance of the tumor's SUV characteristics. The metrics for similarity are SSIM \cite{renieblas2017structural} and RMSE:

\begin{equation}
\label{rmse}
 \text{RMSE} =  \sqrt{\frac{1}{n}\sum _{i=1}^{n}(Y_{i} - \hat {Y_{i}})^{2}}
\end{equation}

The SSIM parameters in our study are the same as in scikit-image library. In this report, we used ISSIM=1-SSIM instead of SSIM as ISSIM is more convenient for similar images, and SSIM is higher than 0.9 for most original and denoised PET. We defined relative metrics in the same way as in \cite{sanaat2020projection}:
\begin{equation}
\label{rmse_rel}
\text{relRMSE} = 1 - \frac {1}{n}\sum_{i=0}^{n} { \frac{\text{RMSE}(denoising(LT_{i}),FT_{i})}{\text{RMSE}(LT_{i},FT_{i})}}
\end{equation}
\eqref{rmse_rel} demonstrates the improvement of the denoising method for noised LT image. The relative ISSIM is defined similarly as \eqref{rmse_rel}. The relative metric changes in range $-\infty$ to 100\%. The negative value means that the method has deteriorated the quality of the image, 0\% - there are no changes, 100\% - the image has been fully denoised and coincides with the original one.

The relative ISSIM in Table \ref{table:rmse_ssim_30sec} are the mean value for the relative ISSIM for each PET image pair and, therefore, could not be obtained from the absolute ISSIM values of Table \ref{table:rmse_ssim_30sec}.

\subsection{SUV error estimation}
\begin{figure}[!t]
\centering
\includegraphics[width=2.2in]{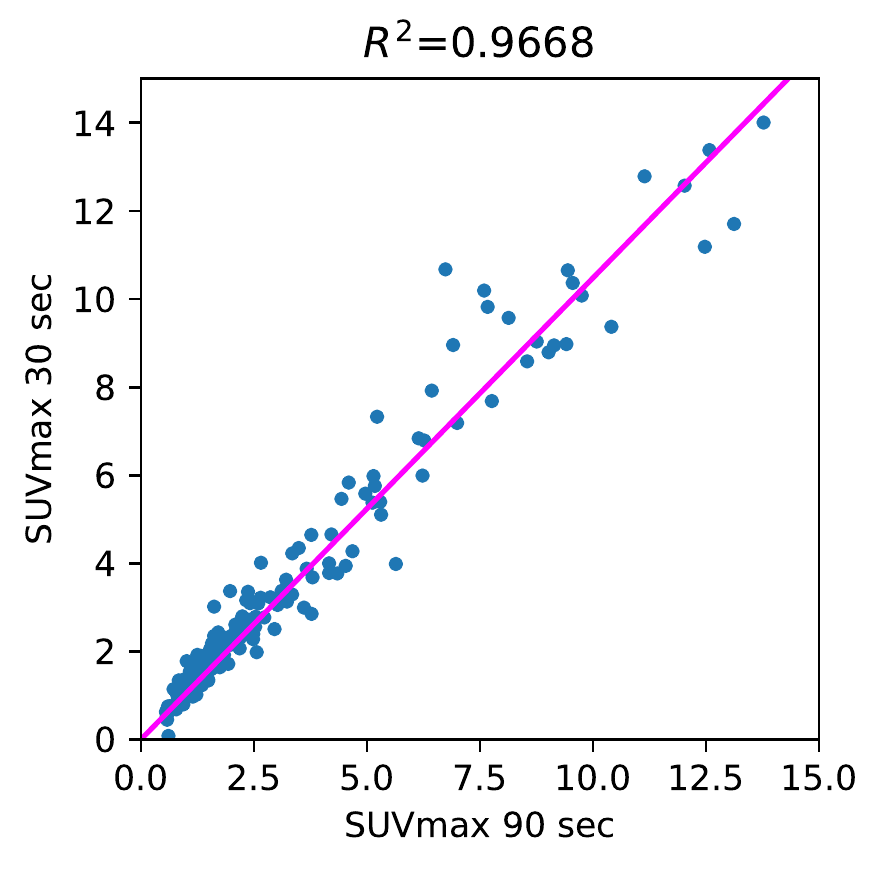}
\caption{Correlation between 90 sec and 30 sec PET SUVmax (regions of interests)}
\label{fig_r2_90_30}
\end{figure}

The use of standard uptake values (SUV) is now commonplace \cite{fletcher2010pet} in clinical FDG-PET/CT oncology imaging and has a specific role in assessing patient response to cancer therapy. SUVmean, SUVpeak \cite{sher2016avid}, and SUVmax are the values commonly used in PET studies.
There are many ways to estimate the correlation between pairs of SUV values for the FT original PET and denoised PET reconstructed from LT. The most common are bias and STD in terms of Bland-Alman plots \cite{weyts2022artificial, katsari2021artificial, lei2020low} and R2 (Fig. \ref{fig_r2_90_30}).

CT is available along with the ground truth PET and could enhance the quality of the tumor segmentation. Instead of employing renginologist for the malignant tumor detection, we segmented tumors automatically in 3D  with the help of nnUnet \cite{isensee2021nnu}. The pretrained weights are the same as in the AutoPET competition baseline \cite{gatidis2022whole}. The nnUnet neural network manipulated two channels (PET \& CT) input with 400$\times$400 resolution. The CT and PET images are to be resized as they have 512$\times$512, and 256$\times$256 resolution.

\begin{figure*}[!t]
\centering
\includegraphics[width=5.0in]{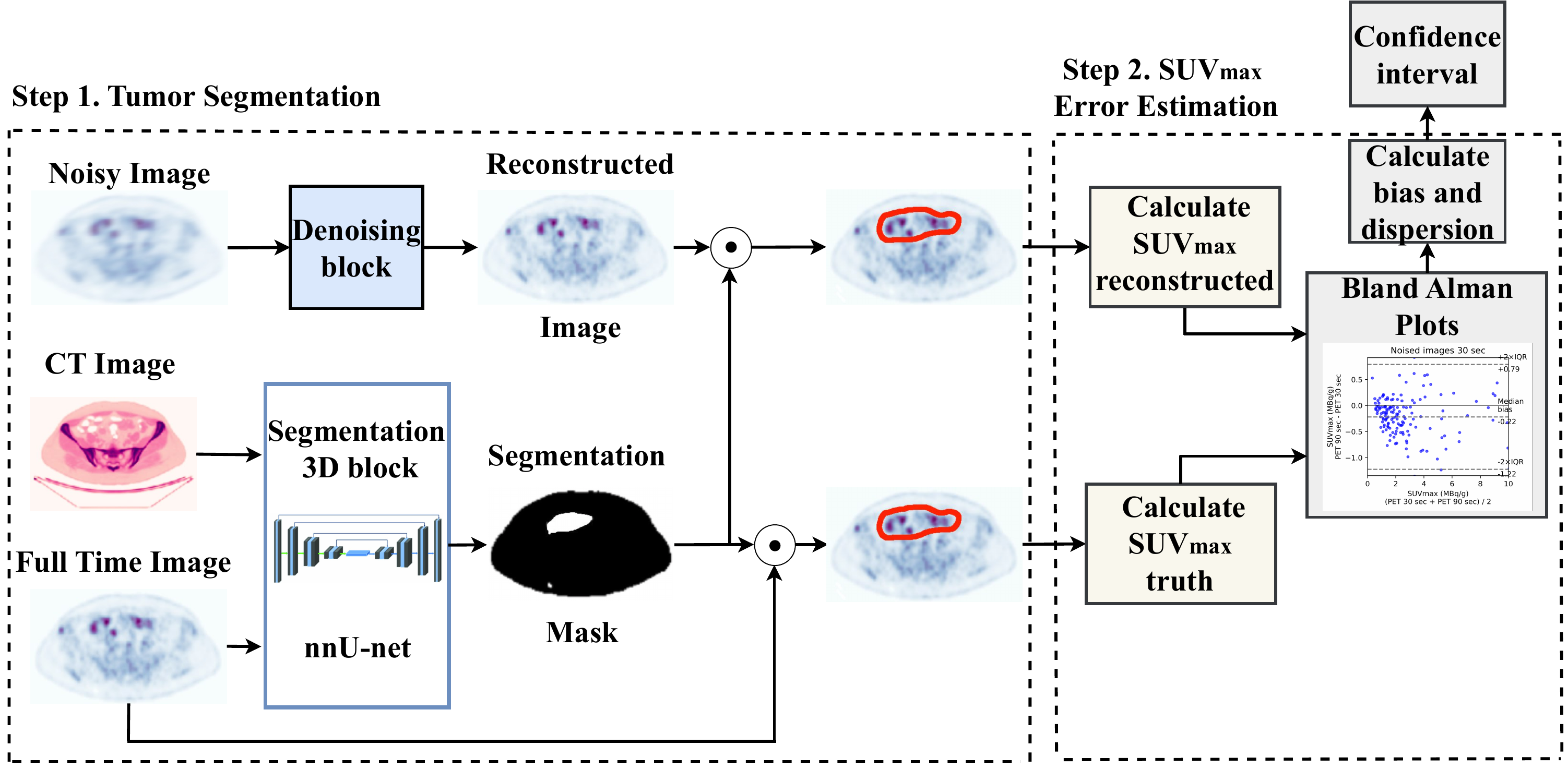}%
\hfil
\caption{SUV confidence interval estimation scheme with automatic tumor segmentation}
\label{fig_segment_scheme}
\end{figure*} 

Fig. \ref{fig_segment_scheme} illustrates the SUV confidence interval estimation scheme. After the nnUnet segmentation cc3d library extracts 3D connected components and separates different tumors. We excluded from the study tumors with a maximum length of less than 7 mm and an average SUV of less than 0.5. Bland-Alman plot is a standard instrument of data similarity evaluation in biomedical research. The plot operates with the region of interest SUV for original and denoised PET. The Bland-Alman plot's bias and dispersion are indicators of denoising quality and are used in the latest step of scheme in Fig. \ref{fig_segment_scheme} for the confidence interval assessment.

The number of tumors selected in the test data is 74. More is needed to get reliable statistics on SUV errors. That is why we decided to utilize validation data for SUV error estimation. This approach is correct as we maximized SSIM during hyper-parameter optimization on the validation data and did not consider SUV, and SUV error for validation and split data have similar values. ISSIM and RMSE metrics were estimated using test split only. The validation data contains 97 tumors; therefore, the total number of tumors in validation and test data is 171.

\section{Neural network implementation and training details}
\label{section:neural_network}
\subsection{Convolutional networks}
Unet and ResNet models, pix2pix and CycleGAN are based on the pytorch implementation of CycleGAN\footnote{\url{https://github.com/junyanz/pytorch-CycleGAN-and-pix2pix}}.

The number of channels in the bottleneck for both models is 64. The Unet model has 54.4 mil. parameters, the ResNet served as encoder has 11.4 mil. parameters themself, and the decoder has 0.37 mil parameters - 11.8 mil parameters in total. The decoder exploits transposed convolutions and does not have skip connections.

Pix2Pix and CycleGAN models use PatchGAN \cite{isola2017image} with 2.8 mil. parameters. Table \ref{table:model_parameters} demonstrates the total number of parameters and the models' size. CUDA memory usage is the maximum size of GPU memory required by the model, depending on the batch size. We used pytorch memory allocated function with SGD optimizer for the memory size estimation, as Adam uses twice as much memory. CUDA memory column in the table \ref{table:model_parameters} does not include GPU memory required for the pytorch framework.

\begin{table}[!t]
\caption{Model parameters.}
\label{table:model_parameters}
\centering
\begin{tabular}{|c||c||c|}
\hline
\thead{Method} & \thead{parameters \\ mil.} &\thead{CUDA \\memory \\Mb} \\
\hline \makecell{SwinIR} & 12.5 &131 + 1300*batch\\
\hline \makecell{ResNet} &11.36 &47 + 341*batch\\
\hline \makecell{Unet} &54.4 &210 + 35*batch\\
\hline \makecell{CycleGAN \\ ResNet} &28.25 &113 + 1370*batch\\
\hline \makecell{CycleGAN \\ UNet} &114.3 &440 + 140*batch \\
\hline
\end{tabular}
\end{table} 

ResNet, Unet, and CycleGAN models predict the difference between noised and denoised images. The SwinIR model has this difference built into its architecture. The Pix2Pix GAN discriminator also used image difference to distinguish between noised and denoised PET. This simple approach applied for the PET denoising improved the results significantly but was used before only in the transformer-based model for CT denoising

L1 loss is used in all models (except for unsupervised CycleGAN) to optimize the similarity between denoised LT and FT images. Pix2pix GAN also uses Euclidean adversarial loss.

\begin{figure*}[!b]
\centering
\includegraphics[width=7.0in]{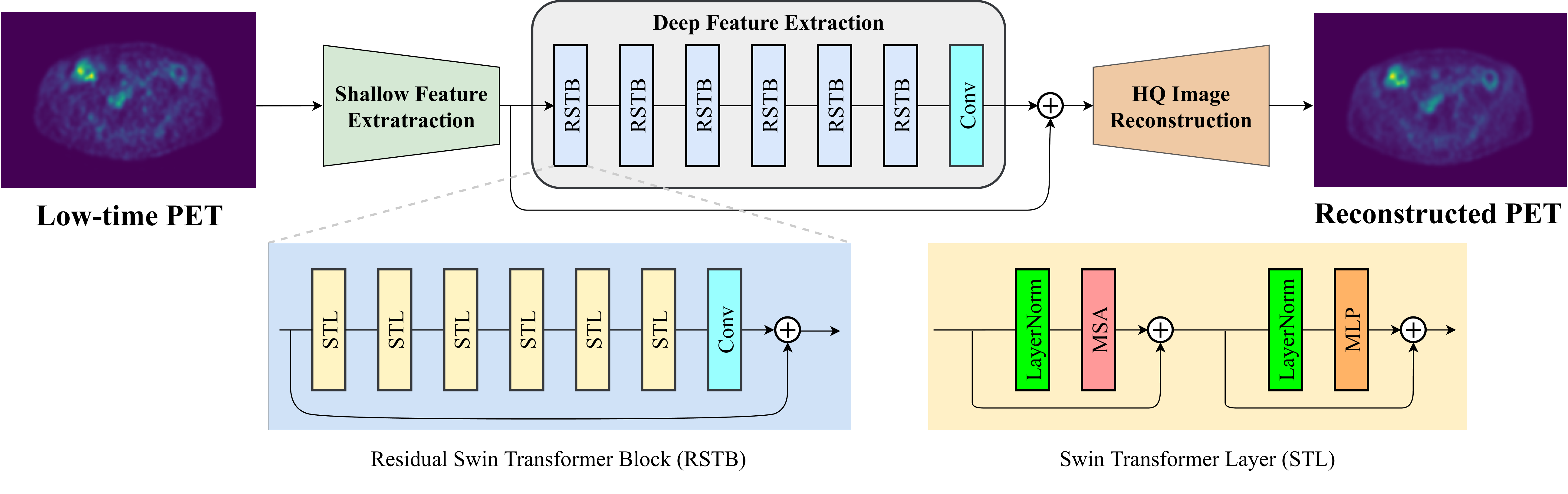}%
\hfil
\caption{The architecture of the SwinIR \cite{liang2021swinir}}
\label{fig_SwinIR}
\end{figure*} 

\subsection{SwinIR}
SwinIR \cite{liang2021swinir} integrates the advantages of both CNN and transformer. On the one hand, it has the advantage of CNN to process images with a large size due to the local attention mechanism. On the other hand, it has the benefit of the transformer to model long-range dependency with the shifted window \cite{liu2021swin}. SwinIR exceeded state-of-the-art CNN performance in denoising and JPEG compression artifact reduction. We implemented code from official SwinIR repository\footnote{\url{https://github.com/JingyunLiang/SwinIR}}.

SwinIR consists of three modules (Fig. \ref{fig_SwinIR}) shallow feature extraction, deep feature extraction, and high-quality image reconstruction modules. The shallow feature extraction module uses a convolution layer to extract shallow features directly transmitted to the reconstruction module to preserve low-frequency information.

The deep feature extraction module is mainly composed of residual Swin Transformer blocks (RSTB in fig. \ref{fig_SwinIR}), each of which utilizes several Swin Transformer layers for local attention and cross-window interaction. In addition, \cite{liang2021swinir} added a convolution layer at the end of the block for feature enhancement and used a residual connection. In the end, shallow and deep features are fused in the reconstruction module for high-quality image reconstruction. The patch size of SwinIR in our training is 64; the window size is 8.

\subsection{Training details}
The hyperparameters tuning is done on the validation data set by maximizing SSIM with optuna library. SSIM is preferrable over L1 and RMSE as it, more than other metrics, coincides with human perception and makes denoised PET look similar to the original \cite{renieblas2017structural}.

We considered identity and image prior loss coefficients between 0 and 30 and weight decay for Unet in the 0.001 - 0.2. The ISSIM dependence of image prior loss coefficient looks the same as in \cite{tang2019unpaired}. The quality of denoising is stable to the identity loss coefficient but could deteriorate up to 20\% of its value when choosing a coefficient higher than 18. 

The augmentations used in training are 360 deg rotation, reflection, and random corps. The augmentations did not improve metrics significantly for ResNet, but they made the training process more stable. In contrast, CycleGAN with ResNet backbone metrics slightly dropped when trained with augmented images. The Unet performance improved significantly after applying augmentations but still left behind ResNet; this could be partly due to overfitting, as Unet has more parameters than ResNet.

Adam is an optimizer for the training process. Unet was trained with weight decay=0.002 to prevent overfitting. That improved relative ISSIM from 27.8\% to 29.0\%. The usage of dropout has a similar effect. The learning rate was chosen individually to achieve the best performance for each model. We trained supervised methods and pix2pix GANs with ResNet backbones using cos learning rate schedule, max lr=0.0002 for 35 epochs.

CycleGAN training includes 30 epochs with a constant learning rate 0.0001, linearly reduced to zero for the following 15 epochs. Optuna library helped to fit the optimal learning rate schedule for SwinIR. Table \ref{table:training_parameters} presents the time required for training each model with one GPU Tesla V100 with except of SwinIR trained on Tesla A100.

\begin{table}[!t]
\caption{Training parameters.}
\label{table:training_parameters}
\centering
\begin{tabular}{|c||c||c||c||c||c|}
\hline
\thead{Method} & \thead{epoch \\ number} &\thead{Time \\h} &\thead{Max \\Lr}&\thead{Lr \\ schedule}&\thead{Batch \\ size}\\
\hline \makecell{SwinIR} & 80 &48 &0.00023 &\makecell{Reduce on\\Plateu} &32\\
\hline \makecell{ResNet} & 35 &9 &0.0002 &Cos &32\\
\hline \makecell{Unet} & 35 &3 &0.0002 &Cos &32\\
\hline \makecell{\\ Pix2pix \\ GAN \\ResNet} & 35 &9 &0.0002 &Cos &32\\
\hline \makecell{CycleGAN \\ ResNet} & 50 &46.5 &0.0001 &Linear &16\\
\hline
\end{tabular}
\end{table} 
 
We trained models with batch size 32 except CycleGAN.The original CycleGAN \cite{zhu2017unpaired} used batch size 1. Unlike the original work in the recent study \cite{lupion2022using} batch size that generates the best PSNR value is 64, using the initial learning rate. The experiments demonstrated that the batch size does not have to be 1 or 4, but it depends on the size of the data set and the type of problem. Therefore, we trained CycleGAN with batch size 16.

\section{Results}
\label{section_results}
The prediction of the difference between noised and denoised images rather than the prediction of the denoised image itself significantly (twice as much for ISSIM metric) improved the quality of ResNet, Unet, and CycleGAN denoising as it is easy for the network to produce noise rather than images \cite{chen2018image}. Our study revealed that 2.5 Gaussian convolution improves RMSE at the cost of SSIM. 

\begin{table}[!t]
\caption{Similarity of 2D PET images - original 90 sec and reconstructed from 30 sec}
\label{table:rmse_ssim_30sec}
\centering
\begin{tabular}{|c||c||c||c|}
\hline Model & \multicolumn{1}{c}{Absolute values} & \multicolumn{2}{c|}{Relative values} \\
\hline
&\thead{1-SSIM} &\thead{RMSE} &\thead{1-SSIM}\\
\hline \makecell{30 vs 90} &8.58$\times$10$\textsuperscript{-2}$ &0\% &0\%\\
\hline \makecell{Gauss conv}  &7.17$\times$10$\textsuperscript{-2}$ &6.8\% &15.3\%\\
\hline \makecell{SwinIR}  &6.06$_{\pm0.04}$$\times$10$\textsuperscript{-2}$ &22.9$_{\pm0.28}$\%  &29.7$_{\pm0.50}$\%\\
\hline \makecell{ResNet}  &$\mathbf{5.77}_{\pm0.06}$$\times$10$\textsuperscript{-2}$ &$\mathbf{24.4}_{\pm0.59}$\% &$\mathbf{32.9}_{\pm0.82}$\%\\
\hline \makecell{UNet}  &6.09$_{\pm0.04}$$\times$10$\textsuperscript{-2}$ &22.0$_{\pm0.46}$\% &29.1$_{\pm0.32}$\%\\
\hline \makecell{Cycle\\GAN sup}  &6.14$_{\pm0.12}$$\times$10$\textsuperscript{-2}$  &20.1$_{\pm1.0}$\% &28.1$_{\pm1.0}$\% \\
\hline \makecell{CycleGAN} &6.29$_{\pm0.12}$$\times$10$\textsuperscript{-2}$ &19.2$_{\pm1.2}$\% &26.6$_{\pm1.3}$\%\\
\hline \makecell{CycleGAN \\ identity}  &6.22$_{\pm0.17}$$\times$10$\textsuperscript{-2}$ & 20.4$_{\pm1.5}$\% &27.3$_{\pm2.2}$\%\\
\hline \makecell{CycleGAN \\ image prior}  &6.24$_{\pm0.07}$$\times$10$\textsuperscript{-2}$ &19.1$_{\pm0.6}$\% &27.1$_{\pm0.78}$\%\\
\hline
\end{tabular}
\end{table}

\begin{figure*}[!t]
\centering
\subfloat[]{\includegraphics[width=1.8in]{original.pdf}%
\label{fig2_first_case}}
\hfil
\subfloat[]{\includegraphics[width=1.8in]{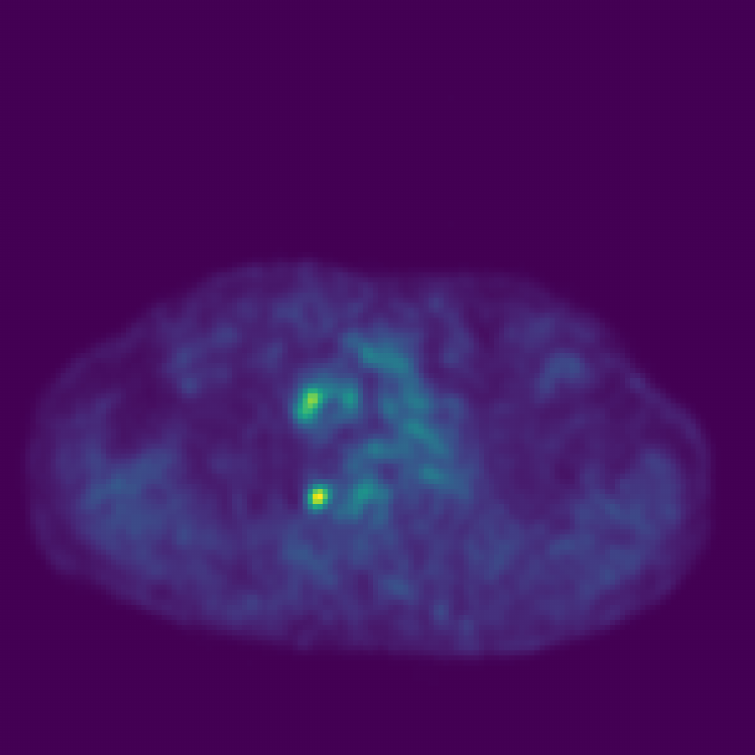}%
\label{fig2_second_case}}
\hfil
\subfloat[]{\includegraphics[width=1.8in]{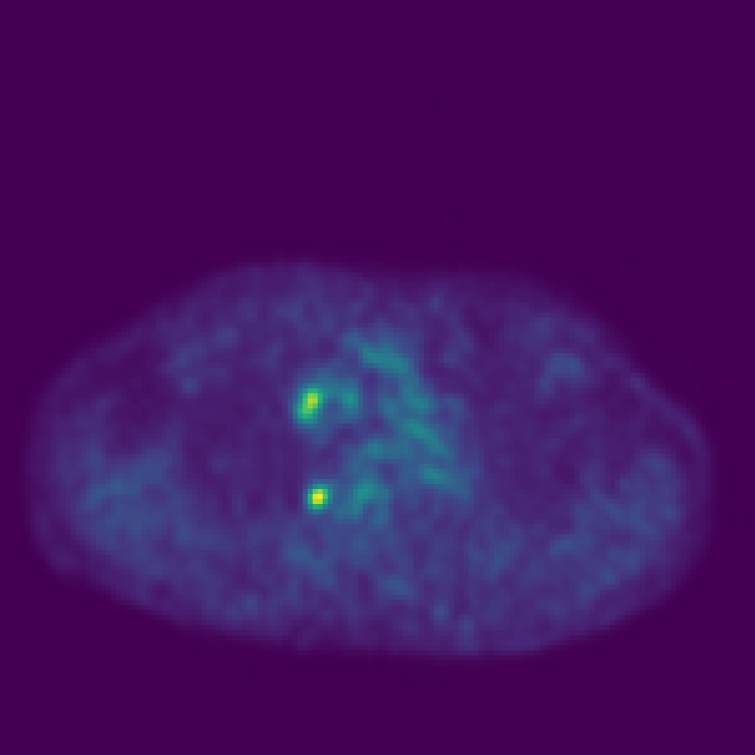}%
\label{fig2_third_case}}
\caption{Low-time PET reconstruction. (a) Full-time PET (90 sec). (b) Reconstructed PET, CycleGAN. (c) Reconstructed PET, Pix2Pix GAN}
\label{fig_example2}
\end{figure*}

\begin{table}[!t]
\caption{SUVpeak characteristics original. PET reconstructed from 30 sec with respect to the original 90 sec PET.}
\label{table:suv_peak_30sec}
\centering
\begin{tabular}{|c||c||c||c|}
\hline
\thead{Method} & \thead{1 - R$\textsuperscript{2}$} &\thead{Median \\bias} &\thead{IQR}\\
\hline \makecell{30 vs 90} &8.48$\times$10$\textsuperscript{-3}$ &1.52$\times$10$\textsuperscript{-2}$ &0.1262\\
\hline \makecell{Gauss convol} &9.53$\times$10$\textsuperscript{-3}$ &$\mathbf{0.85}\times$10$\textsuperscript{-2}$ &0.1112\\
\hline \makecell{SwinIR} &9.2$_{\pm0.7}$$\times$10$\textsuperscript{-3}$ &2.6$_{\pm0.6}$$\times$10$\textsuperscript{-2}$ &$\mathbf{0.099}_{\pm0.006}$\\
\hline \makecell{ResNet} &8.8$_{\pm1.9}$$\times$10$\textsuperscript{-3}$ &1.7$_{\pm1.1}$$\times$10$\textsuperscript{-2}$ &0.107$_{\pm0.006}$\\
\hline \makecell{Unet} &9.0$_{\pm1.6}$$\times$10$\textsuperscript{-3}$ &2.0$_{\pm2.4}$$\times$10$\textsuperscript{-2}$ &0.103$_{\pm0.009}$\\
\hline \makecell{Cycle\\GAN sup} &8.0$_{\pm0.5}$$\times$10$\textsuperscript{-3}$ &1.8$_{\pm0.5}$$\times$10$\textsuperscript{-2}$ &0.106$_{\pm0.003}$ \\
\hline \makecell{CycleGAN} & 7.6$_{\pm0.25}$$\times$10$\textsuperscript{-3}$ &1.3$_{\pm1.2}$$\times$10$\textsuperscript{-2}$ &0.109$_{\pm0.006}$\\
\hline \makecell{CycleGAN \\ identity} & $\mathbf{7.5}_{\pm0.23}\times$10$\textsuperscript{-3}$ &0.9$_{\pm0.8}$$\times$10$\textsuperscript{-2}$ &0.107$_{\pm0.009}$ \\
\hline \makecell{CycleGAN \\ image prior} &8.0$_{\pm0.25}$$\times$10$\textsuperscript{-3}$  &2.1$_{\pm0.7}$$\times$10$\textsuperscript{-2}$ &0.113$_{\pm0.003}$ \\
\hline
\end{tabular}
\end{table} 

\begin{table}[!t]
\caption{SUVmax characteristics original. PET reconstructed from 30 sec with respect to the original 90 sec PET.}
\label{table:suv_max_30sec}
\centering
\begin{tabular}{|c||c||c||c|}
\hline
\thead{Method} & \thead{1 - R$\textsuperscript{2}$} &\thead{Median \\bias} &\thead{IQR}\\
\hline \makecell{30 vs 90} & 3.061$\times$10$\textsuperscript{-2}$ &-21.78$\times$10$\textsuperscript{-2}$ &0.5031\\
\hline \makecell{Gauss convol} & 8.282$\times$10$\textsuperscript{-2}$ &9.71$\times$10$\textsuperscript{-2}$ &0.4667\\
\hline \makecell{SwinIR} &$\mathbf{1.8}_{\pm0.19}$$\times$10$\textsuperscript{-2}$  &18.1$_{\pm2.6}$$\times$10$\textsuperscript{-2}$ &$\mathbf{0.34}_{\pm0.03}$\\
\hline \makecell{ResNet} &2.1$_{\pm0.5}$$\times$10$\textsuperscript{-2}$ &11.0$_{\pm12.0}$$\times$10$\textsuperscript{-2}$ &0.35$_{\pm0.09}$ \\
\hline \makecell{Unet} &19.6$_{\pm4.4}$$\times$10$\textsuperscript{-2}$  &13.1$_{\pm1.9}$$\times$10$\textsuperscript{-2}$ &0.38$_{\pm0.07}$ \\
\hline \makecell{Cycle\\GAN sup} &2.1$_{\pm0.21}$$\times$10$\textsuperscript{-2}$ &6.2$_{\pm4.0}$$\times$10$\textsuperscript{-2}$ &$\mathbf{0.34}_{\pm0.028}$ \\
\hline \makecell{CycleGAN} &2.1$_{\pm0.37}$$\times$10$\textsuperscript{-2}$ &2.4$_{\pm1.6}$$\times$10$\textsuperscript{-2}$ &0.38$_{\pm0.05}$  \\
\hline \makecell{CycleGAN \\ identity} &1.96$_{\pm0.12}$$\times$10$\textsuperscript{-2}$  &5.4$_{\pm4.9}$$\times$10$\textsuperscript{-2}$  &0.36$_{\pm0.04}$  \\
\hline \makecell{CycleGAN \\ image prior} &2.12$_{\pm0.27}$$\times$10$\textsuperscript{-2}$ &$\mathbf{1.4}_{\pm2.9}$$\times$10$\textsuperscript{-2}$ &0.39$_{\pm0.06}$ \\
\hline
\end{tabular}
\end{table} 
 
\begin{table}[!t]
\caption{Confidence interval for median SUVpeak and SUVmax error (MBq/kG). PET reconstructed from 30 sec with respect to the original 90 sec PET.}
\label{table:conf_inter_suv_30}
\centering
\begin{tabular}{|c||c||c||c||c|}
\hline Model & \multicolumn{2}{c}{SUVpeak} & \multicolumn{2}{c|}{SUVmax} \\
\hline
&\thead{Lower} & \thead{Upper} &\thead{Lower} &\thead{Upper}\\
\hline \makecell{30 vs 90} &-0.212 &0.242 &-1.123 &0.687\\
\hline \makecell{ResNet} &-0.197 &0.231 &-0.802 &1.022\\
\hline \makecell{SwinIR} &-0.169 &0.221 &-0.511 &0.873\\
\hline \makecell{CycleGAN \\ sup} &-0.183 &$\mathbf{0.219}$&-0.640 &$\mathbf{0.764}$\\
\hline
\end{tabular}
\end{table} 
 
The ISSIM and RMSE metrics are in Table \ref{table:rmse_ssim_30sec}. ResNet backbone with convolutional decoder without skip-connection outperformed Unet and SwinIR in all cases, that is why in Table \ref{table:rmse_ssim_30sec} and the following tables, we presented the results for CycleGAN and Pix2pix GAN for the ResNet backbone only. All networks were trained independently five times to estimate the confidence intervals (95\% confidence level).

ResNet has the best image similarity metrics among all methods for both weakly noised (60 sec) and strongly noised (30 sec) PET. SwinIR and Unet follow it. The quality of SwinIR PET restoration is better than that of Unet. At the same time \cite{jang2022spach} demonstrates better Swin performance over Unet on 25\% low-count data. This fact indicates that convolutional layers of SwinIR for shallow feature extraction improve reconstruction quality compared to the pure Swin transformer architecture.

Pix2Pix GAN (ResNet + PatchGAN discriminator) without distance loss (Fig. \ref{fig2_third_case}) produces realistic denoised images but with low metrics. The SSIM steadily improves while the distance loss coefficient increases and reaches its plateau when the GAN degrades to a simple supervised model, as distance loss outweighs adversarial losses. Our Pix2Pix GAN did not show a higher quality over other methods as 3D CVT-GAN in \cite{zeng20223d}, or BiC-GAN \cite{fei2022classification} for the brain's PET synthesis. 

As the original Pix2Pix paper \cite{isola2017image} mentioned, the random input $z$ does not change the GAN results, and the model is indeed deterministic. We concluded that the Pix2Pix GAN model is not appropriate for the denoising problem as the adversarial loss improves image appearance rather than SSIM. So it produces a realistic but far from the original images. For that reason, we have not included pix2pix GAN metrics in Tables \ref{table:rmse_ssim_30sec} \ref{table:suv_peak_30sec}, \ref{table:suv_max_30sec}.
We trained PixPix GAN with the coefficient by distance loss 10.0 as it does not outweigh the adversarial loss. 

\begin{figure}[!t]
\centering
\includegraphics[width=2.5in]{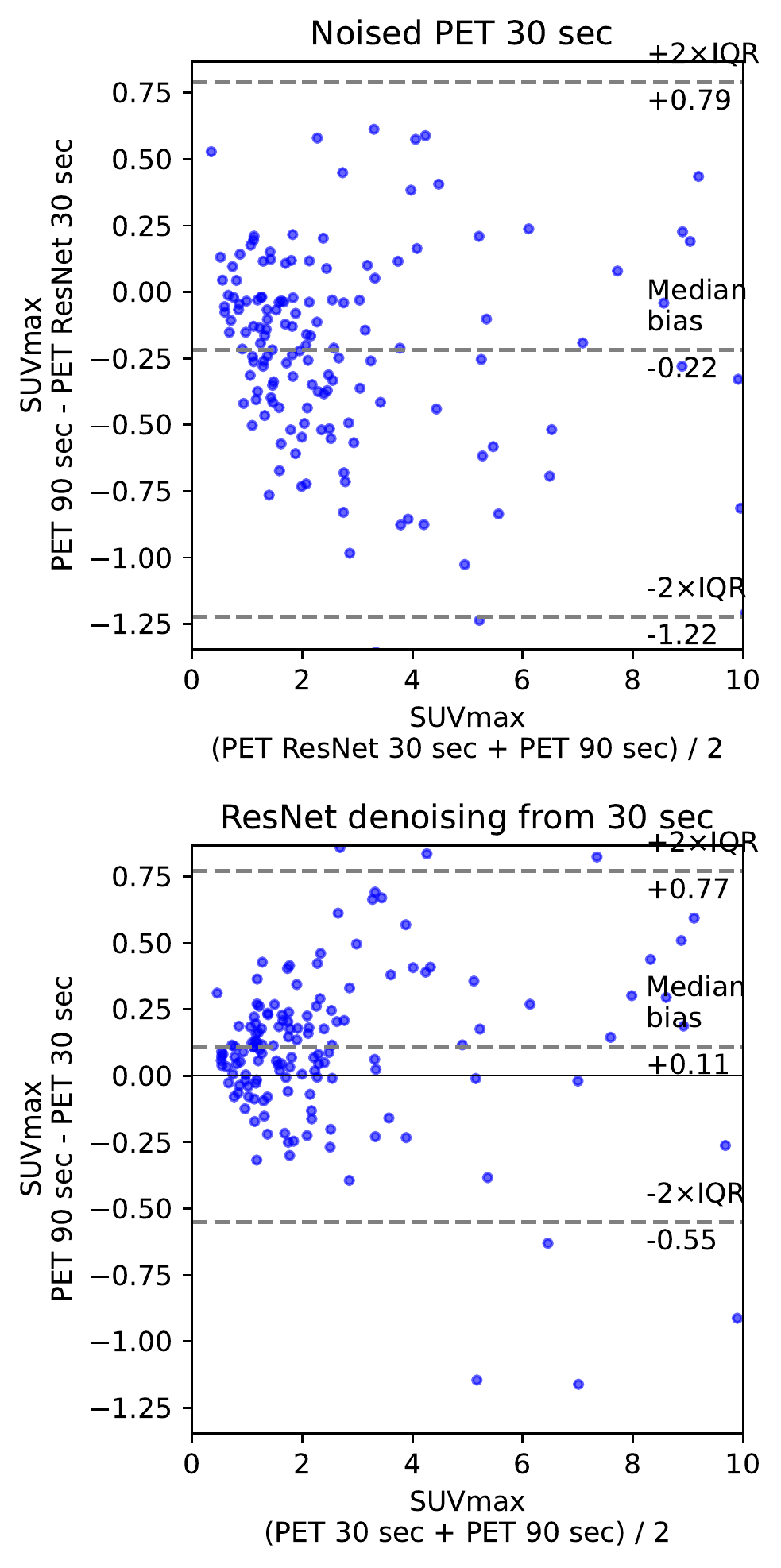}
\caption{Bland-Alman plot for 30 sec PET before and after denoising}
\label{fig_bland_alman_30}
\end{figure}

Unsupervised CycleGAN provides weaker image similarity metrics in comparison with supervised. The optimal coefficient for the identity loss is 2.2, and the image prior loss is 9.2. For the distance loss in supervised CycleGAN we used the same coefficient as for image prior loss - 9.2. 

For the reduced 30\% data set, the impact of the image prior loss is the same as in \cite{tang2019unpaired}. But for the full-size data set image prior loss effect is less pronounced. The reason is that for the small size data set image prior loss works as regularization and prevents overfitting. For example, the size of the data set in \cite{tang2019unpaired} is only 906 CT images with 512$\times$512 resolution

The image prior loss improves the quality and stability of the reconstruction. The identity loss has a more profound effect on the SSIM and RMSE metrics than the image prior loss for the weakly noised PET \ref{table:rmse_ssim_60sec}. We used the same coefficient for image prior loss as for the distance loss in supervised CycleGAN - 9.2. Supervised CycleGAN image similarity metrics lie between supervised and unsupervised methods but have the advantage of CycleGAN estimating SUVmax with lower bias and dispersion. 

Our results contradict \cite{lei2019whole} where CycleGAN outperformed supervised Unet and Unet GAN. That could stem from the small size of the data set used in \cite{lei2019whole}, or the reason is the usage in CycleGAN backbone other than Unet.

Tables \ref{table:suv_peak_30sec}, \ref{table:suv_max_30sec} represent discrepancy (Median bias and IQR values for the Bland-Alman plot) in SUVpeak and SUVmax estimation. We preferred median and IQR over mean and STD like \cite{weyts2022artificial} did, as these metrics are more robust to the outliers. 

ResNet possesses the best image similarity metrics in all cases, whereas no method outperforms others in SUV estimation for all criteria and parameters as Tables \ref{table:suv_peak_30sec}, \ref{table:suv_max_30sec}, \ref{table:conf_inter_suv_30} show.

Pix2pix ResNet GAN has sub-optimal SUVmax estimation metrics. Table \ref{table:conf_inter_suv_30} represents the confidence interval of SUVpeak and SUVmax error median bias $\pm$ 1.8*IQR. The SUVmax error confidence interval for ResNet denoised data is (-0.802, 1.005) vs. (-1.226, 0.786) for unprocessed 30 sec PET. CycleGAN provides the lowest among maximal absolute values of the SUVmax confidence interval for 30 sec PET reconstruction. These values are highlighted in Tables \ref{table:conf_inter_suv_30}, \ref{table:conf_inter_suv_60} because they are essential for the renginologists as SUVmax is critical for cancer diagnosis and treatment and. 

The SUVpeak/max confidence intervals for 60 sec. PET are (-0.107, 0.106), (-0.47, 0.44).

The supervised methods produce reconstructed PET with positive SUVmax bias (Fig. \ref{fig_bland_alman_30} ) because they flatten the signals too much. On the other hand, CycleGAN family methods have predominantly negative or around zero bias like SubtlePET algorithm \cite{weyts2022artificial}, which is an advantage of unsupervised methods.

The SSIM metric was an optimization goal for the network, which is why the model having the lowest ISSIM and RMSE results does not necessarily produce the best SUV reconstruction for tumors. In \cite{sanaei2021does} the SUVmean bias was not improved by HighResNet (LD contains only 6\% of the FD) even though the PSNR and SSIM of the reconstructed image were better than the LD. There is also a high variation in SUV between different training of the same network.

\section{Clinical discussion}
\label{section_clinical_discussion}
Due to biological or technological factors, SUV may significantly differ from one measurement to another \cite{adams2010systematic, lindholm2014repeatability, de2022method}. For example, technological factors include inter-scanner variability, image reconstruction, processing parameter changes, and calibration error between a scanner and a dose calibrator.

An example of biological factors are respiratory motion \cite{guo2022unsupervised} like cardiac motion, and body motion. Patient's breathing can  affect measured SUVs, particularly in lung bases or upper abdomen lesions. As \cite{adams2010systematic} mentioned "This occurs because CT (used for attenuation correction during PET image reconstruction) can occur during a single breath-hold of the patient, but a PET acquisition for a given bed position takes minutes and is obtained while the patient is quietly breathing. If the diaphragm position in CT does not match the average position during PET, the attenuation correction may over- or undercorrect the radioactivity concentration, which would change the measured SUV." The machine learning methods \cite{guo2022mcp, zeng2022supervised} try fixing the problem.

The studies \cite{schwartz2011repeatability, lodge2012noise, burger2012repeatability} reported high $\Delta$SUVmax between two FDG acquisitions. One should consider the details of these experiments to compare them with the results of our study.

The recent work \cite{de2022method} conducted the following experiment. The six phantom spheres of 10 - 37 mm diameters were filled with the concentration 20.04 MBq/ml. The FT 150-s mode was divided into subsets of shorter frames varying from 4 to 30 sec. The SUVmax monotonically increases with sphere's diameter. The ratio of the standard SUVmax deviation to its average value for 30 sec PET is about 15\% for a 10 mm sphere and the confidence interval length achieves up to 0.5 kBq/ml. The experiment \cite{de2022method} does not take into account biological and most of technical factors \cite{adams2010systematic}, so the final SUVmax discrepancy between two PET could achieve higher values.

The aforementioned estimation of SUVmax discrepancy for the same tumor between two PET acquisitions shows that SUVmax denoising error for the 30 and 60 sec PET achieved in our study lies in the acceptable range.

\section{Conclusion}
PET denoising may allow for reducing an injected dose or increasing the scanner’s throughput. We reconstructed PET with a reduced acquisition time 30 and 60 sec and compared it with the original full-time 90 sec PET. We trained and tested SwinIR, Unet, ResNet, and CycleGAN with ResNet backbone and different auxiliary losses for that purpose. 

The supervised denoising methods have significantly better RMSE and ISSIM than unsupervised ones. This result differs from previous studies claiming that CycleGAN surpasses Unet and ResNet. The ResNet reconstructs PET images with the lowest RMSE and ISSIM outperforming SwinIR and Unet. Supervised CycleGAN achieved the lowest SUVmax error after PET denoising. The SUVmax error of the reconstructed PET is comparable with the reproducibility error due to biological or technological factors.

It is a matter of discussion on which metric - SUVmax error or visual similarity should be given priority. Image similarity metrics provide visual information that can help doctors determine if a tumor is malignant or benign. Image comparison also helps clinicians assess treatment efficacy by comparing pre-treatment images with post-treatment ones to measure any changes due to therapy intervention. On the other hand, SUVmax error provides quantitative data regarding how much a tumor has reduced in size after treatment interventions have been applied; this allows physicians an objective way of evaluating treatments' effectiveness without relying solely on subjective visual assessments from image comparisons alone.

The future work is to find a better combination of the supervised and unsupervised methods to achieve the lowest SUV error while preserving the high SSIM of the enhanced PET.


\bibliography{sample}

\begin{thebibliography}{10}
\providecommand{\url}[1]{#1}
\csname url@samestyle\endcsname
\providecommand{\newblock}{\relax}
\providecommand{\bibinfo}[2]{#2}
\providecommand{\BIBentrySTDinterwordspacing}{\spaceskip=0pt\relax}
\providecommand{\BIBentryALTinterwordstretchfactor}{4}
\providecommand{\BIBentryALTinterwordspacing}{\spaceskip=\fontdimen2\font plus
\BIBentryALTinterwordstretchfactor\fontdimen3\font minus
  \fontdimen4\font\relax}
\providecommand{\BIBforeignlanguage}[2]{{%
\expandafter\ifx\csname l@#1\endcsname\relax
\typeout{** WARNING: IEEEtran.bst: No hyphenation pattern has been}%
\typeout{** loaded for the language `#1'. Using the pattern for}%
\typeout{** the default language instead.}%
\else
\language=\csname l@#1\endcsname
\fi
#2}}
\providecommand{\BIBdecl}{\relax}
\BIBdecl

\bibitem{schaefferkoetter2019low}
J.~Schaefferkoetter, Y.-H. Nai, A.~Reilhac, D.~W. Townsend, L.~Eriksson, and
  M.~Conti, ``Low dose positron emission tomography emulation from decimated
  high statistics: a clinical validation study,'' \emph{Medical physics},
  vol.~46, no.~6, pp. 2638--2645, 2019.

\bibitem{yang2022quasi}
G.~Yang, C.~Li, Y.~Yao, G.~Wang, and Y.~Teng, ``Quasi-supervised learning for
  super-resolution pet,'' \emph{arXiv preprint arXiv:2209.01325}, 2022.

\bibitem{jang2022spach}
S.-I. Jang, T.~Pan, Y.~Li, P.~Heidari, J.~Chen, Q.~Li, and K.~Gong, ``Spach
  transformer: Spatial and channel-wise transformer based on local and global
  self-attentions for {PET} image denoising,'' \emph{arXiv preprint
  arXiv:2209.03300}, 2022.

\bibitem{hu2022transem}
R.~Hu and H.~Liu, ``{TransEM}: Residual swin-transformer based regularized
  {PET} image reconstruction,'' in \emph{Medical Image Computing and Computer
  Assisted Intervention--MICCAI 2022: 25th International Conference, Singapore,
  September 18--22, 2022, Proceedings, Part IV}.\hskip 1em plus 0.5em minus
  0.4em\relax Springer, 2022, pp. 184--193.

\bibitem{spuhler2020full}
K.~Spuhler, M.~Serrano-Sosa, R.~Cattell, C.~DeLorenzo, and C.~Huang,
  ``Full-count {PET} recovery from low-count image using a dilated
  convolutional neural network,'' \emph{Medical Physics}, vol.~47, no.~10, pp.
  4928--4938, 2020.

\bibitem{liang2021swinir}
J.~Liang, J.~Cao, G.~Sun, K.~Zhang, L.~Van~Gool, and R.~Timofte, ``{SwinIR}:
  Image restoration using swin transformer,'' in \emph{Proceedings of the
  IEEE/CVF International Conference on Computer Vision}, 2021, pp. 1833--1844.

\bibitem{lu2019investigation}
W.~Lu, J.~A. Onofrey, Y.~Lu, L.~Shi, T.~Ma, Y.~Liu, and C.~Liu, ``An
  investigation of quantitative accuracy for deep learning based denoising in
  oncological pet,'' \emph{Physics in Medicine \& Biology}, vol.~64, no.~16, p.
  165019, 2019.

\bibitem{sanaat2021deep}
A.~Sanaat, I.~Shiri, H.~Arabi, I.~Mainta, R.~Nkoulou, and H.~Zaidi, ``Deep
  learning-assisted ultra-fast/low-dose whole-body {PET/CT} imaging,''
  \emph{European journal of nuclear medicine and molecular imaging}, vol.~48,
  no.~8, pp. 2405--2415, 2021.

\bibitem{weyts2022artificial}
K.~Weyts, C.~Lasnon, R.~Ciappuccini, J.~Lequesne, A.~Corroyer-Dulmont, E.~Quak,
  B.~Clarisse, L.~Roussel, S.~Bardet, and C.~Jaudet, ``Artificial
  intelligence-based pet denoising could allow a two-fold reduction in [18f]
  {FDG PET} acquisition time in digital {PET/CT},'' \emph{European Journal of
  Nuclear Medicine and Molecular Imaging}, pp. 1--11, 2022.

\bibitem{bonardel2022clinical}
G.~Bonardel, A.~Dupont, P.~Decazes, M.~Queneau, R.~Modzelewski, J.~Coulot,
  N.~Le~Calvez, and S.~Hapdey, ``Clinical and phantom validation of a deep
  learning based denoising algorithm for {F-18-FDG PET} images from lower
  detection counting in comparison with the standard acquisition,''
  \emph{EJNMMI physics}, vol.~9, no.~1, pp. 1--23, 2022.

\bibitem{sanaat2020projection}
A.~Sanaat, H.~Arabi, I.~Mainta, V.~Garibotto, and H.~Zaidi, ``Projection space
  implementation of deep learning--guided low-dose brain {PET} imaging improves
  performance over implementation in image space,'' \emph{Journal of Nuclear
  Medicine}, vol.~61, no.~9, pp. 1388--1396, 2020.

\bibitem{katsari2021artificial}
K.~Katsari, D.~Penna, V.~Arena, G.~Polverari, A.~Ianniello, D.~Italiano,
  R.~Milani, A.~Roncacci, R.~O. Illing, and E.~Pelosi, ``Artificial
  intelligence for reduced dose {18F-FDG} {PET} examinations: a real-world
  deployment through a standardized framework and business case assessment,''
  \emph{EJNMMI physics}, vol.~8, no.~1, pp. 1--15, 2021.

\bibitem{chen2022image}
K.~T. Chen and G.~Zaharchuk, ``Image synthesis for low-count {PET}
  acquisitions: lower dose, shorter time,'' in \emph{Biomedical Image Synthesis
  and Simulation}.\hskip 1em plus 0.5em minus 0.4em\relax Elsevier, 2022, pp.
  369--391.

\bibitem{liu2021artificial}
J.~Liu, M.~Malekzadeh, N.~Mirian, T.-A. Song, C.~Liu, and J.~Dutta,
  ``Artificial intelligence-based image enhancement in pet imaging: Noise
  reduction and resolution enhancement,'' \emph{PET clinics}, vol.~16, no.~4,
  pp. 553--576, 2021.

\bibitem{gong2018pet}
K.~Gong, J.~Guan, C.-C. Liu, and J.~Qi, ``{PET} image denoising using a deep
  neural network through fine tuning,'' \emph{IEEE Transactions on Radiation
  and Plasma Medical Sciences}, vol.~3, no.~2, pp. 153--161, 2018.

\bibitem{sanaei2021does}
B.~Sanaei, R.~Faghihi, H.~Arabi, and H.~Zaidi, ``Does prior knowledge in the
  form of multiple low-dose {PET} images (at different dose levels) improve
  standard-dose {PET} prediction?'' in \emph{2021 IEEE Nuclear Science
  Symposium and Medical Imaging Conference (NSS/MIC)}.\hskip 1em plus 0.5em
  minus 0.4em\relax IEEE, 2021, pp. 1--3.

\bibitem{lei2019whole}
Y.~Lei, X.~Dong, T.~Wang, K.~Higgins, T.~Liu, W.~J. Curran, H.~Mao, J.~A. Nye,
  and X.~Yang, ``Whole-body {PET} estimation from low count statistics using
  cycle-consistent generative adversarial networks,'' \emph{Physics in Medicine
  \& Biology}, vol.~64, no.~21, p. 215017, 2019.

\bibitem{cui2019pet}
J.~Cui, K.~Gong, N.~Guo, C.~Wu, X.~Meng, K.~Kim, K.~Zheng, Z.~Wu, L.~Fu, B.~Xu
  \emph{et~al.}, ``{PET} image denoising using unsupervised deep learning,''
  \emph{European journal of nuclear medicine and molecular imaging}, vol.~46,
  pp. 2780--2789, 2019.

\bibitem{manakov2019noise}
I.~Manakov, M.~Rohm, C.~Kern, B.~Schworm, K.~Kortuem, and V.~Tresp, ``Noise as
  domain shift: Denoising medical images by unpaired image translation,'' in
  \emph{Domain adaptation and representation transfer and medical image
  learning with less labels and imperfect data}.\hskip 1em plus 0.5em minus
  0.4em\relax Springer, 2019, pp. 3--10.

\bibitem{kwon2021cycle}
T.~Kwon and J.~C. Ye, ``Cycle-free cyclegan using invertible generator for
  unsupervised low-dose {CT} denoising,'' \emph{IEEE Transactions on
  Computational Imaging}, vol.~7, pp. 1354--1368, 2021.

\bibitem{tang2019unpaired}
C.~Tang, J.~Li, L.~Wang, Z.~Li, L.~Jiang, A.~Cai, W.~Zhang, N.~Liang, L.~Li,
  and B.~Yan, ``Unpaired low-dose {CT} denoising network based on
  cycle-consistent generative adversarial network with prior image
  information,'' \emph{Computational and mathematical methods in medicine},
  vol. 2019, 2019.

\bibitem{lei2020low}
Y.~Lei, T.~Wang, X.~Dong, K.~Higgins, T.~Liu, W.~J. Curran, H.~Mao, J.~A. Nye,
  and X.~Yang, ``Low dose {PET} imaging with {CT}-aided cycle-consistent
  adversarial networks,'' in \emph{Medical Imaging 2020: Physics of Medical
  Imaging}, vol. 11312.\hskip 1em plus 0.5em minus 0.4em\relax SPIE, 2020, pp.
  1043--1049.

\bibitem{chandrashekar2022deep}
A.~Chandrashekar, A.~Handa, J.~Ward, V.~Grau, and R.~Lee, ``A deep learning
  pipeline to simulate fluorodeoxyglucose ({FDG}) uptake in head and neck
  cancers using non-contrast {CT} images without the administration of
  radioactive tracer,'' \emph{Insights into imaging}, vol.~13, no.~1, pp.
  1--10, 2022.

\bibitem{li2021novel}
Y.~Li, K.~Zhang, W.~Shi, Y.~Miao, and Z.~Jiang, ``A novel medical image
  denoising method based on conditional generative adversarial network,''
  \emph{Computational and Mathematical Methods in Medicine}, vol. 2021, 2021.

\bibitem{sano2021denoising}
A.~Sano, T.~Nishio, T.~Masuda, and K.~Karasawa, ``Denoising {PET} images for
  proton therapy using a residual {U-net},'' \emph{Biomedical Physics \&
  Engineering Express}, vol.~7, no.~2, p. 025014, 2021.

\bibitem{schaefferkoetter2020convolutional}
J.~Schaefferkoetter, J.~Yan, C.~Ortega, A.~Sertic, E.~Lechtman, Y.~Eshet,
  U.~Metser, and P.~Veit-Haibach, ``Convolutional neural networks for improving
  image quality with noisy {PET} data,'' \emph{EJNMMI research}, vol.~10, pp.
  1--11, 2020.

\bibitem{Luthra2021EformerEE}
A.~Luthra, H.~Sulakhe, T.~Mittal, A.~Iyer, and S.~K. Yadav, ``Eformer: Edge
  enhancement based transformer for medical image denoising,'' \emph{ArXiv},
  vol. abs/2109.08044, 2021.

\bibitem{cui2022pet}
J.~Cui, Y.~Xie, A.~A. Joshi, K.~Gong, K.~Kim, Y.-D. Son, J.-H. Kim, R.~Leahy,
  H.~Liu, and Q.~Li, ``{PET} denoising and uncertainty estimation based on
  {NVAE} model using quantile regression loss,'' in \emph{Medical Image
  Computing and Computer Assisted Intervention--MICCAI 2022: 25th International
  Conference, Singapore, September 18--22, 2022, Proceedings, Part IV}.\hskip
  1em plus 0.5em minus 0.4em\relax Springer, 2022, pp. 173--183.

\bibitem{zhu2017unpaired}
J.-Y. Zhu, T.~Park, P.~Isola, and A.~A. Efros, ``Unpaired image-to-image
  translation using cycle-consistent adversarial networks,'' in
  \emph{Proceedings of the IEEE international conference on computer vision},
  2017, pp. 2223--2232.

\bibitem{park2021effect}
K.~S. Park, S.-G. Cho, J.~Kim, and H.-C. Song, ``The effect of weights for
  cycle-consistency loss and identity loss on blood-pool image to bone image
  translation with {CycleGAN},'' 2021.

\bibitem{renieblas2017structural}
G.~P. Renieblas, A.~T. Nogu{\'e}s, A.~M. Gonz{\'a}lez, N.~G. Le{\'o}n, and
  E.~G. Del~Castillo, ``Structural similarity index family for image quality
  assessment in radiological images,'' \emph{Journal of medical imaging},
  vol.~4, no.~3, p. 035501, 2017.

\bibitem{fletcher2010pet}
J.~Fletcher and P.~Kinahan, ``{PET/CT} standardized uptake values ({SUVs}) in
  clinical practice and assessing response to therapy,'' \emph{NIH Public
  Access}, vol.~31, no.~6, pp. 496--505, 2010.

\bibitem{sher2016avid}
A.~Sher, F.~Lacoeuille, P.~Fosse, L.~Vervueren, A.~Cahouet-Vannier, D.~Dabli,
  F.~Bouchet, and O.~Couturier, ``For avid glucose tumors, the suv peak is the
  most reliable parameter for {[18F] FDG-PET/CT} quantification, regardless of
  acquisition time,'' \emph{EJNMMI research}, vol.~6, no.~1, pp. 1--6, 2016.

\bibitem{isensee2021nnu}
F.~Isensee, P.~F. Jaeger, S.~A. Kohl, J.~Petersen, and K.~H. Maier-Hein,
  ``{nnU-Net}: a self-configuring method for deep learning-based biomedical
  image segmentation,'' \emph{Nature methods}, vol.~18, no.~2, pp. 203--211,
  2021.

\bibitem{gatidis2022whole}
S.~Gatidis, T.~Hepp, M.~Fr{\"u}h, C.~La~Foug{\`e}re, K.~Nikolaou,
  C.~Pfannenberg, B.~Sch{\"o}lkopf, T.~K{\"u}stner, C.~Cyran, and D.~Rubin, ``A
  whole-body {FDG-PET/CT} dataset with manually annotated tumor lesions,''
  \emph{Scientific Data}, vol.~9, no.~1, pp. 1--7, 2022.

\bibitem{isola2017image}
P.~Isola, J.-Y. Zhu, T.~Zhou, and A.~A. Efros, ``Image-to-image translation
  with conditional adversarial networks,'' in \emph{Proceedings of the IEEE
  conference on computer vision and pattern recognition}, 2017, pp. 1125--1134.

\bibitem{liu2021swin}
Z.~Liu, Y.~Lin, Y.~Cao, H.~Hu, Y.~Wei, Z.~Zhang, S.~Lin, and B.~Guo, ``Swin
  transformer: Hierarchical vision transformer using shifted windows,'' in
  \emph{Proceedings of the IEEE/CVF International Conference on Computer
  Vision}, 2021, pp. 10\,012--10\,022.

\bibitem{lupion2022using}
M.~Lupi{\'o}n, J.~Sanjuan, and P.~Ortigosa, ``Using a multi-gpu node to
  accelerate the training of pix2pix neural networks,'' \emph{The Journal of
  Supercomputing}, pp. 1--18, 2022.

\bibitem{chen2018image}
J.~Chen, J.~Chen, H.~Chao, and M.~Yang, ``Image blind denoising with generative
  adversarial network based noise modeling,'' in \emph{Proceedings of the IEEE
  conference on computer vision and pattern recognition}, 2018, pp. 3155--3164.

\bibitem{zeng20223d}
P.~Zeng, L.~Zhou, C.~Zu, X.~Zeng, Z.~Jiao, X.~Wu, J.~Zhou, D.~Shen, and
  Y.~Wang, ``{3D CVT-GAN}: A {3D} convolutional vision transformer-gan for
  {PET} reconstruction,'' in \emph{Medical Image Computing and Computer
  Assisted Intervention--MICCAI 2022: 25th International Conference, Singapore,
  September 18--22, 2022, Proceedings, Part VI}.\hskip 1em plus 0.5em minus
  0.4em\relax Springer, 2022, pp. 516--526.

\bibitem{fei2022classification}
Y.~Fei, C.~Zu, Z.~Jiao, X.~Wu, J.~Zhou, D.~Shen, and Y.~Wang,
  ``Classification-aided high-quality {PET} image synthesis via bidirectional
  contrastive gan with shared information maximization,'' in \emph{Medical
  Image Computing and Computer Assisted Intervention--MICCAI 2022: 25th
  International Conference, Singapore, September 18--22, 2022, Proceedings,
  Part VI}.\hskip 1em plus 0.5em minus 0.4em\relax Springer, 2022, pp.
  527--537.

\bibitem{adams2010systematic}
M.~C. Adams, T.~G. Turkington, J.~M. Wilson, and T.~Z. Wong, ``A systematic
  review of the factors affecting accuracy of {SUV} measurements,''
  \emph{American Journal of Roentgenology}, vol. 195, no.~2, pp. 310--320,
  2010.

\bibitem{lindholm2014repeatability}
H.~Lindholm, J.~Staaf, H.~Jacobsson, F.~Brolin, R.~Hatherly, and
  A.~S{\^a}nchez-Crespo, ``Repeatability of the maximum standard uptake value
  ({SUVmax}) in {FDG PET},'' \emph{Molecular imaging and radionuclide therapy},
  vol.~23, no.~1, p.~16, 2014.

\bibitem{de2022method}
G.~M. De~Luca and J.~B. Habraken, ``Method to determine the statistical
  technical variability of suv metrics,'' \emph{EJNMMI physics}, vol.~9, no.~1,
  p.~40, 2022.

\bibitem{guo2022unsupervised}
X.~Guo, B.~Zhou, D.~Pigg, B.~Spottiswoode, M.~E. Casey, C.~Liu, and N.~C.
  Dvornek, ``Unsupervised inter-frame motion correction for whole-body dynamic
  {PET} using convolutional long short-term memory in a convolutional neural
  network,'' \emph{Medical Image Analysis}, vol.~80, p. 102524, 2022.

\bibitem{guo2022mcp}
X.~Guo, B.~Zhou, X.~Chen, C.~Liu, and N.~C. Dvornek, ``{MCP-Net}: Inter-frame
  motion correction with patlak regularization for whole-body dynamic {PET},''
  in \emph{Medical Image Computing and Computer Assisted Intervention--MICCAI
  2022: 25th International Conference, Singapore, September 18--22, 2022,
  Proceedings, Part IV}.\hskip 1em plus 0.5em minus 0.4em\relax Springer, 2022,
  pp. 163--172.

\bibitem{zeng2022supervised}
T.~Zeng, J.~Zhang, E.~Revilla, E.~V. Lieffrig, X.~Fang, Y.~Lu, and J.~A.
  Onofrey, ``Supervised deep learning for head motion correction in {PET},'' in
  \emph{Medical Image Computing and Computer Assisted Intervention--MICCAI
  2022: 25th International Conference, Singapore, September 18--22, 2022,
  Proceedings, Part IV}.\hskip 1em plus 0.5em minus 0.4em\relax Springer, 2022,
  pp. 194--203.

\bibitem{schwartz2011repeatability}
J.~Schwartz, J.~Humm, M.~Gonen, H.~Kalaigian, H.~Schoder, S.~Larson, and
  S.~Nehmeh, ``Repeatability of {SUV} measurements in serial {PET},''
  \emph{Medical physics}, vol.~38, no.~5, pp. 2629--2638, 2011.

\bibitem{lodge2012noise}
M.~A. Lodge, M.~A. Chaudhry, and R.~L. Wahl, ``Noise considerations for {PET}
  quantification using maximum and peak standardized uptake value,''
  \emph{Journal of Nuclear Medicine}, vol.~53, no.~7, pp. 1041--1047, 2012.

\bibitem{burger2012repeatability}
I.~A. Burger, D.~M. Huser, C.~Burger, G.~K. von Schulthess, and A.~Buck,
  ``Repeatability of {FDG} quantification in tumor imaging: averaged {SUVs} are
  superior to {SUVmax},'' \emph{Nuclear medicine and biology}, vol.~39, no.~5,
  pp. 666--670, 2012.

\end{thebibliography}
\bibliographystyle{IEEEtran}

{\appendix
The appendix contains models metrics for the 60 sec PET reconstruction not included in the main body of the article.  

\begin{table}[h]
\caption{Similarity of 2D PET images - original 90 sec and reconstructed from 60 sec. ResNet backbone}
\label{table:rmse_ssim_60sec}
\centering
\begin{tabular}{|c||c||c||c|}
\hline Model & \multicolumn{1}{c}{Absolute values} & \multicolumn{2}{c|}{Relative values} \\
\hline
 & \thead{1-SSIM} &\thead{RMSE} &\thead{1-SSIM}\\
\hline
 &1-SSIM &RMSE &1-SSIM \\
\hline \makecell{30 vs 90}  &2.61$\times$10$\textsuperscript{-2}$ &0\% &0\%\\
\hline \makecell{Gauss conv}  &2.44$\times$10$\textsuperscript{-2}$ &3.21\% &5.3\%\\
\hline \makecell{SwinIR} &2.27$_{\pm0.011}$$\times$10$\textsuperscript{-2}$ &9.0$_{\pm0.27}$\% &12.7$_{\pm0.5}$\%\\
\hline \makecell{ResNet} &$\mathbf{2.24}_{\pm0.016}$$\times$10$\textsuperscript{-2}$ &$\mathbf{9.3}_{\pm0.42}$\% &$\mathbf{13.8}_{\pm0.65}$\%\\
\hline \makecell{UNet} &2.29$_{\pm0.015}$$\times$10$\textsuperscript{-2}$ &8.4$_{\pm0.28}$\% &12.4$_{\pm0.53}$\%\\
\hline \makecell{CycleGAN} &2.33$_{\pm0.011}$$\times$10$\textsuperscript{-2}$ &6.6$_{\pm0.27}$\% &10.3$_{\pm0.23}$\%\\
\hline \makecell{Cycle\\GAN sup} &2.32$_{\pm0.018}$$\times$10$\textsuperscript{-2}$ &6.8$_{\pm0.44}$\% &10.8$_{\pm0.84}$\%\\
\hline \makecell{CycleGAN \\ identity} &2.32$_{\pm0.011}$$\times$10$\textsuperscript{-2}$ &7.1$_{\pm0.54}$\% &11.0$_{\pm0.27}$\%\\
\hline \makecell{CycleGAN \\ image prior} &2.33$_{\pm0.013}$$\times$10$\textsuperscript{-2}$ &6.8$_{\pm0.51}$\% &10.7$_{\pm0.49}$\%\\
\hline
\end{tabular}
\end{table}

\begin{table}[h]
\caption{SUVpeak characteristics original. PET reconstructed from 60 sec with respect to the original 90 sec PET. ResNet backbone. Median values }
\label{table:suv_peak_60sec}
\centering
\begin{tabular}{|c||c||c||c|}
\hline
\thead{Method} & \thead{1 - R$\textsuperscript{2}$} &\thead{Median \\bias} &\thead{IQR}\\
\hline \makecell{60 vs 90} &2.19$\times$10$\textsuperscript{-3}$ &-0.2$\times$10$\textsuperscript{-3}$ &5.92$\times$10$\textsuperscript{-2}$ \\
\hline \makecell{Gauss convol} &2.2$\times$10$\textsuperscript{-3}$ &1.3$\times$10$\textsuperscript{-3}$ &6.46$\times$10$\textsuperscript{-2}$ \\
\hline \makecell{SwinIR} &2.4$_{\pm0.11}$$\times$10$\textsuperscript{-3}$ &6.6$_{\pm2.5}$$\times$10$\textsuperscript{-3}$ &5.6$_{\pm0.27}$$\times$10$\textsuperscript{-2}$ \\
\hline \makecell{UNet} &$\mathbf{2.0}_{\pm0.15}$$\times$10$\textsuperscript{-3}$  &3.0$_{\pm9.0}$$\times$10$\textsuperscript{-3}$ &$\mathbf{5.4}_{\pm0.7}$$\times$10$\textsuperscript{-2}$ \\
\hline \makecell{ResNet} &2.1$_{\pm0.021}$$\times$10$\textsuperscript{-3}$ &5.36$_{\pm4.9}$$\times$10$\textsuperscript{-3}$ &5.6$_{\pm0.6}$$\times$10$\textsuperscript{-2}$ \\
\hline \makecell{Cycle\\GAN sup} &2.1$_{\pm0.05}$$\times$10$\textsuperscript{-3}$  &$\mathbf{-0.06}_{\pm3.4}$$\times$10$\textsuperscript{-3}$ &5.6$_{\pm0.4}$$\times$10$\textsuperscript{-2}$  \\
\hline \makecell{CycleGAN} &2.07$_{\pm0.04}$$\times$10$\textsuperscript{-3}$ &0.3$_{\pm2.3}$$\times$10$\textsuperscript{-3}$ &5.4$_{\pm0.4}$$\times$10$\textsuperscript{-2}$ \\
\hline \makecell{CycleGAN \\ identity} &2.07$_{\pm0.018}$$\times$10$\textsuperscript{-3}$  &-2.8$_{\pm2.1}$$\times$10$\textsuperscript{-3}$ &5.46$_{\pm0.18}$$\times$10$\textsuperscript{-2}$ \\
\hline \makecell{CycleGAN \\ image prior} &2.09$_{\pm0.018}$$\times$10$\textsuperscript{-3}$  &-1.2$_{\pm3.3}$$\times$10$\textsuperscript{-3}$ &5.81$_{\pm0.24}$$\times$10$\textsuperscript{-2}$ \\
\hline
\end{tabular}
\end{table}

\begin{table}[!t]
\caption{SUVmax characteristics original. PET reconstructed from 60 sec with respect to the original 90 sec PET. ResNet backbone. \label{tab:table4}}
\label{table:suv_max_60sec}
\centering
\begin{tabular}{|c||c||c||c|}
\hline
\thead{Method} & \thead{1 - R$\textsuperscript{2}$} &\thead{Median \\bias} &\thead{IQR}\\
\hline \makecell{60 vs 90} &5.93$\times$10$\textsuperscript{-3}$  &-5.95$\times$10$\textsuperscript{-2}$ &0.2489 \\
\hline \makecell{Gauss convol} &16.4$\times$10$\textsuperscript{-3}$  &3.01$\times$10$\textsuperscript{-2}$ &0.2545 \\
\hline \makecell{SwinIR} &$\mathbf{3.8}_{\pm0.47}$$\times$10$\textsuperscript{-3}$  &6.6$_{\pm1.3}$$\times$10$\textsuperscript{-2}$ &0.247$_{\pm0.048}$ \\
\hline \makecell{ResNet} &5.3$_{\pm0.25}$$\times$10$\textsuperscript{-3}$   &2.9$_{\pm2.7}$$\times$10$\textsuperscript{-2}$ &$\mathbf{0.216}_{\pm0.029}$ \\
\hline \makecell{UNet} &5.4$_{\pm0.4}$$\times$10$\textsuperscript{-3}$   &2.9$_{\pm1.5}$$\times$10$\textsuperscript{-2}$ &0.241$_{\pm0.032}$ \\
\hline \makecell{Cycle\\GAN sup} &5.5$_{\pm0.3}$$\times$10$\textsuperscript{-3}$   &-1.3$_{\pm0.9}$$\times$10$\textsuperscript{-2}$ &0.241$_{\pm0,008}$ \\
\hline \makecell{CycleGAN} &5.7$_{\pm0.4}$$\times$10$\textsuperscript{-3}$   &-1.0$_{\pm1.2}$$\times$10$\textsuperscript{-2}$ &0.249$_{\pm0.021}$ \\
\hline \makecell{CycleGAN \\ identity} &5.6$_{\pm0.4}$$\times$10$\textsuperscript{-3}$  &$\mathbf{-0.3}_{\pm1.0}$$\times$10$\textsuperscript{-2}$ &0.245$_{\pm0.009}$ \\
\hline \makecell{CycleGAN \\ image prior} &5.3$_{\pm0.13}$$\times$10$\textsuperscript{-3}$  &-0.7$_{\pm1.1}$$\times$10$\textsuperscript{-2}$ &0.242$_{\pm0.017}$  \\
\hline
\end{tabular}
\end{table} 

\begin{table}[!t]
\caption{Confidence interval for the median SUVpeak and SUVmax error. PET reconstructed from 60 sec with respect to the original 90 sec PET.}
\label{table:conf_inter_suv_60}
\centering
\begin{tabular}{|c||c||c||c||c|}
\hline Model & \multicolumn{2}{c}{SUVpeak} & \multicolumn{2}{c|}{SUVmax} \\
\hline
&\thead{Lower} & \thead{Upper} &\thead{Lower} &\thead{Upper}\\
\hline \makecell{60 vs 90} &$\mathbf{-0.1067}$ &0.10636 &-0.5075 &0.3885\\
\hline \makecell{ResNet} &-0.1111 &0.1218 &-0.4372 &0.4952\\
\hline \makecell{SwinIR} &-0.1016 &0.1148 &-0.5146 &0.5553\\
\hline \makecell{CycleGAN \\ sup} &-0.1114 &0.1113 &$\mathbf{-0.4702}$ &0.4442\\
\hline
\end{tabular}
\end{table}

}

\end{document}